\documentclass[]{spie}  

\usepackage{amsmath,amsfonts,amssymb}
\usepackage{graphicx}
\usepackage[colorlinks=true, allcolors=blue]{hyperref}
\usepackage{authblk}
\usepackage{placeins}
\usepackage{array}
\usepackage{subcaption}
\usepackage{caption}
\usepackage[flushleft]{threeparttable}
\usepackage{booktabs}
\DeclareCaptionFont{ninept}{\fontsize{9pt}{11pt}\selectfont}
\usepackage{enumitem}

\newcommand{\mc}[1]{\multicolumn{1}{|l|}{#1}}
\newcommand{\citem}[1]{\hspace{4mm} $\circ$ #1}

\title{High multiplex and precision: the design and development of FLEX, a grid-based fiber positioner with large patrol radius and minimized telecentric error}

\author[a]{Aaron Omadutt}
\author[a]{Roelof S. de Jong}
\author[b]{Will Saunders}
\author[b]{Joseph W. Barrow}
\author[b]{Suryansh Saxena}
\author[b]{Jon Lawrence}
\author[a]{Thomas Liebner}
\author[a]{Frank Dionies}
\affil[a]{Leibniz-Institut für Astrophysik Potsdam, An der Sternwarte 16, 14482 Potsdam, Germany}
\affil[b]{Australian Astronomical Optics, Macquarie University, 2113 Macquarie Park, Sydney, Australia}

\authorinfo{Further author information (send correspondence to): E-mail: aomadutt@aip.de\\}

\begin{document} 
\maketitle

\begin{abstract}
In next-generation spectroscopic facilities, the need for high multiplex fiber positioning systems that can operate within highly constrained focal surface areas is becoming increasingly prevalent (e.g. the Wide-Field Spectroscopic Telescope (WST) with 30,000+ fibers in a 1.4 meter focal surface). The Fiber Location EXtender (FLEX) positioner is being developed to operate within these constraints while performing more efficiently than existing fiber positioners. Specifically, the FLEX positioner focuses on surpassing current astronomical instrument metrics by substantially improving fiber patrol radii while minimizing telecentric error with the telescope pupil and minimizing space between positioners for dense clustering and high Multi-Object Spectrograph (MOS) multiplexing. 
The patented concept of the FLEX positioner utilizes nickel-titanium alloy (Nitinol), a highly flexible metal able to withstand large deflections, in a series of three concentric tubes with differing geometric alterations. The positioner is constructed in a manner that allows its tip to remain inherently parallel with its base as it tilts. Additionally, the internal space within the tubes allows the fiber to run freely along the positioner’s axis minimizing potential Focal Ratio Degradation (FRD). Designed to accommodate a patrol radius of 2.5$\times$ pitch within the WST focal surface architecture, the design delivers a maximum patrol radius of up to $\sim$22.5 mm with a telecentric error of $<0.39^\circ$. FLEX uses three piezoelectric actuators, providing not only a large radial displacement in any direction, but also focus adjustment. To scale this single-positioner architecture to the system level, a modular focal surface layout of 90 identical curvilinear modules has been devised. This layout houses 30240 positioners across a 2-degree hexagonal field-of-view (FoV), accommodating a central hole for the Integral Field pickoff mirror. There are just 3 support struts that cover just 0.8\% of the FoV, and are narrow enough for FLEX to give full coverage across them. 1 in 16 positioners are for High-Resolution, the rest go to 3 sets of Low-Resolution spectrographs; each of these 4 sets of positioners has virtually full coverage of the FoV.

\end{abstract}


\section{INTRODUCTION}
\label{sec:intro} 

The demand for next-generation spectroscopic facilities featuring high-density, high-multiplex fiber positioning systems is rapidly increasing (e.g. WST\cite{Bacon2024}, MegaMapper\cite{Blanc2022}, MUST \cite{Cai_2025}). Modern astrophysics seeks to map the universe across an immense range of physical scales—from the sub-parsec birthplaces of stars to the 100-Megaparsec architecture of the early cosmic web. To address these questions, the Wide-field Spectroscopic Telescope (WST) will conduct an ambitious 18,000 square-degree survey, targeting high-redshift galaxies ($2 < z < 5.5$) to probe an era when the universe was dominated by matter and expansion was decelerating\cite{Bacon2024}. Such a task requires a revolutionary leap in survey speed, necessitating the distribution of over 30,000 independent optical fibers across a 1.4-meter focal surface.

The Fiber Location EXtender (FLEX) positioner was originally proposed as a novel architectural concept in robotic fiber positioning to achieve large patrol radii and high packing densities \cite{DeJong2024}. The system has since been further developed to accommodate the needs of the Wide-field Spectroscopic Telescope. To meet the stringent mechanical and optical requirements of the telescope, which emphasize dense component packing while minimizing throughput loss, the positioner features a small mechanical footprint and intrinsic design features that minimize telecentric errors while maximizing positioning precision.

To contextualize its design, the FLEX positioner can be compared to the proven "tilting spine" architecture pioneered by the Echidna system for FMOS \cite{Gillingham2000}. Tilting spines utilize slender carbon-fiber tubes carrying an optical fiber at the tip to independently patrol target areas. While this established approach allows for exceptionally dense packing and maximizes individual patrol fields, tilting the entire spine inherently causes the fiber face to tilt relative to the telescope's optical focal plane. This introduces both defocus and angular telecentricity errors, both of which degrade overall optical throughput.

In contrast, the patented FLEX architecture departs from the rigid tilting mechanism, utilizing a novel series of nested, geometrically altered Nitinol tubes. This design is engineered to fundamentally minimize tilt-induced telecentricity errors by keeping the fiber tip parallel to the optical axis, while simultaneously incorporating the capability for active focus adjustment.

To contextualize the performance advantages of this concentric tube design relative to prominent robotic fiber positioner technologies currently deployed or proposed for major multi-object spectrograph (MOS) facilities, a comparative overview of key architectural metrics is provided in Table~\ref{tab:positioner_architectures_comparison}.

\begin{table}[ht]
    \centering
    \caption{Comparative evaluation of representative advanced astronomical fiber positioning architectures.}
    \label{tab:positioner_architectures_comparison}
    \small
    \begin{threeparttable}
        \begin{tabular}{@{}lcccccc@{}}
            \toprule
            \textbf{Parameter} & \textbf{FLEX} & \textbf{Echidna}~\cite{Gillingham2000} & \textbf{AESOP}~\cite{deJong2019} & \textbf{LAMOST}~\cite{Xing2010} & \textbf{DESI}~\cite{Silber2023} & \textbf{MOONS}~\cite{Cirami2014} \\
             & (WST) & (FMOS) & (4MOST) & & & \\
            \midrule
            Kinematic Type & Flexure Spine & Tilting Spine & Tilting Spine & Theta-Phi & Theta-Phi & Alpha-Beta \\
            Nominal Pitch & 6.2~/~6.86~mm & 7.0~mm & 9.54~mm & 25.6~mm & 10.4~mm & 25.0~mm \\
            Patrol Radius & $2.5\times$ pitch\textsuperscript{a} & $\sim$7.0~mm & 11.8~mm & 16.5~mm & 6.0~mm & 9--25~mm \\
            Target Multiplex & 30,000+ & 400 & 2,436 & 4,000 & 5,000 & 1,001 \\
            Active $Z$-Focus & Yes & No & No & No & No & No \\
            Positioning Accuracy\textsuperscript{b} & $<15~\mu\text{m}$ & $<10~\mu\text{m}$ & $<10~\mu\text{m}$ & $<40~\mu\text{m}$ & $<5~\mu\text{m}$ & $<20~\mu\text{m}$ \\
            Max Telecentric Error & $\le0.5^\circ$ & $>1.5^\circ$ & $2.64^\circ$ & $<0.5^\circ$ & $<0.26^\circ$ & $<0.25^\circ$ \\
            Collision Complexity\textsuperscript{c} & Low & Low & Low & High & High & High \\
            \bottomrule
        \end{tabular}
        \begin{tablenotes}
            \small
            \item\textsuperscript{a} Corresponds to a physical patrol radius layout of 15.5~mm / 17.15~mm depending on assigned module pitch.
            \item\textsuperscript{b} Design constraint; experimental validation is subject to future prototype testing.
            \item\textsuperscript{c} The collision complexity is defined by the geometric path planning required for a positioner to avoid physical interference with its neighboring units.
        \end{tablenotes}
    \end{threeparttable}
\end{table}
\FloatBarrier

In this paper, we present a technical evaluation of the FLEX positioner’s performance through comprehensive mechanical simulations. We analyze the structural response and positioning accuracy inherent to the system's design, alongside a preliminary focal surface module layout designed to accommodate the 30,000+ positioners required for the WST.

\section{FLEX Design Concept}

The mechanical architecture of the FLEX positioner draws inspiration from millimeter-scale miniature robots utilized in endoscopic surgery, which are designed to navigate complex pathways to transport micro-surgical instruments through the human body. The FLEX system builds upon similar design principles to implement extremely compact concentric tubes that provide exceptional multi-axis maneuverability when actuated. However, a fundamental engineering divergence exists in the actuation constraints. While the physical scale of the structural components remains comparable (at the millimeter level), most medical endoscopic devices rely on external, macroscopic motors capable of generating tens to hundreds of Newtons of force. To house thousands of independent positioners within the minimal footprint required for the WST focal surface, such bulky actuation methods are unfeasible. 

Consequently, the FLEX positioner is engineered to preserve the millimeter-scale form factor and maneuverability of its medical counterparts while optimizing internal mechanics to function with minimal driving force. This optimization allows the system to accommodate ultra-compact micro-actuators while strictly adhering to the spatial allocation of the focal surface.

\subsection{Requirements and Constraints}

To fulfill the requirements of a high-multiplex MOS interface, the design must satisfy both the physical space constraints of a small mounting pitch and the optical requirements for minimizing throughput loss through high-precision positioning. The driving design principles and target specifications for the FLEX system are summarized in Table \ref{tab:technical requirements}. In this specific iteration, the pitch was set at 6.2 mm and 6.86 mm to align with current module design developments for the WST.

\begin{table}[ht]
    \centering
    \caption{FLEX Design Technical Requirements} 
    \label{tab:technical requirements}
    \begin{tabular}{@{}ll@{}} 
        \toprule
        \textbf{Parameter} & \textbf{Requirement} \\ 
        \midrule
        Patrol Radius & $\geq 2.5 \times$ pitch (6.2~mm $|$ 6.86~mm) $\sim$ 15.5~mm $|$ 17.15~mm \\
        Telecentricity Error & $\leq 0.5^{\circ}$ \\
        Positioning Accuracy & $\leq 15~\mu$m (for $\geq 98\%$ active fibers) \\
        Equivalent Stress\textsuperscript{*} & $< 400$~MPa \\
        Actuation Force & $\leq 2.5$~N \\
        Step Resolution & $< 10~\mu$m \\
        Lifetime Reconfigurations & $> 500,000$ cycles \\
        \bottomrule
        \addlinespace
        \multicolumn{2}{@{}l}{
            \begin{minipage}{10cm}
                \small \textsuperscript{*} Based on the Nitinol SE alloy model from the ANSYS Granta Material Data for Simulation (MDS) library\cite{Ansys2026} at 22$^{\circ}$C.
            \end{minipage}
        } \\
    \end{tabular}
\end{table}
\FloatBarrier

The large patrol area and simple collision path avoidance allowed by FLEX provide several significant advantages:

\begin{itemize}
    \item It naturally accommodates the 1:16 ratio in HR/LR fibers required for WST.
    \item It allows targets to be grouped in two or three magnitude ranges such that targets of similar magnitude are going to the same spectrographs, thereby reducing concerns of cross-talk and light scattering in the spectrographs, allowing fibers to be packed closer together in the spectrographs and hence reducing the overall cost for spectrographs and the facility.
    \item It allows for future upgrades for which potentially only a fraction of fibers will be used, for instance to feed near-infrared spectrographs or mini-IFUs (which it accommodates naturally due to its benign strain on the fiber bundle).
    \item It allows for high fiber allocation fractions, even in highly clustered fields, for targets with small angular separation, or in fields where the target density is comparable to the number of fibers.
    \item Its natural low collision cross-section and simple path nature during reconfigurations allow for quick reconfiguration path calculation ``on the fly'', enabling transient insertion up to the last minute or incorporation of the results from the previous exposure into the next (if such results were available already).
\end{itemize}

\subsection{Mechanics - Positioner Body}
\begin{figure} [ht]
    \centering
    \begin{subfigure}[b]{1.0\textwidth}
        \centering
        \includegraphics[width=0.9\textwidth]{"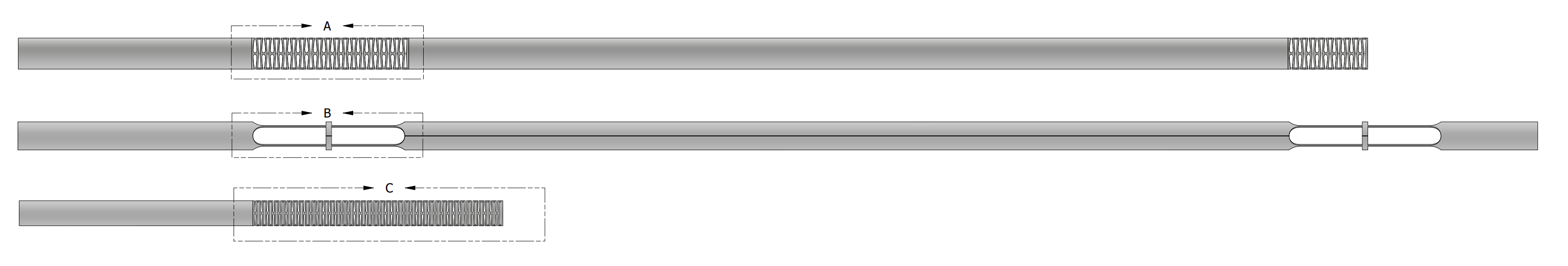"}
        \caption{CAD layout showing individual designs for the outer (OT), middle (MT), and inner (IT) tubes.}
        \label{fig:flex_separated}
    \end{subfigure}

    \begin{subfigure}[b]{1.0\textwidth}
        \centering
        \includegraphics[width=0.85\textwidth]{"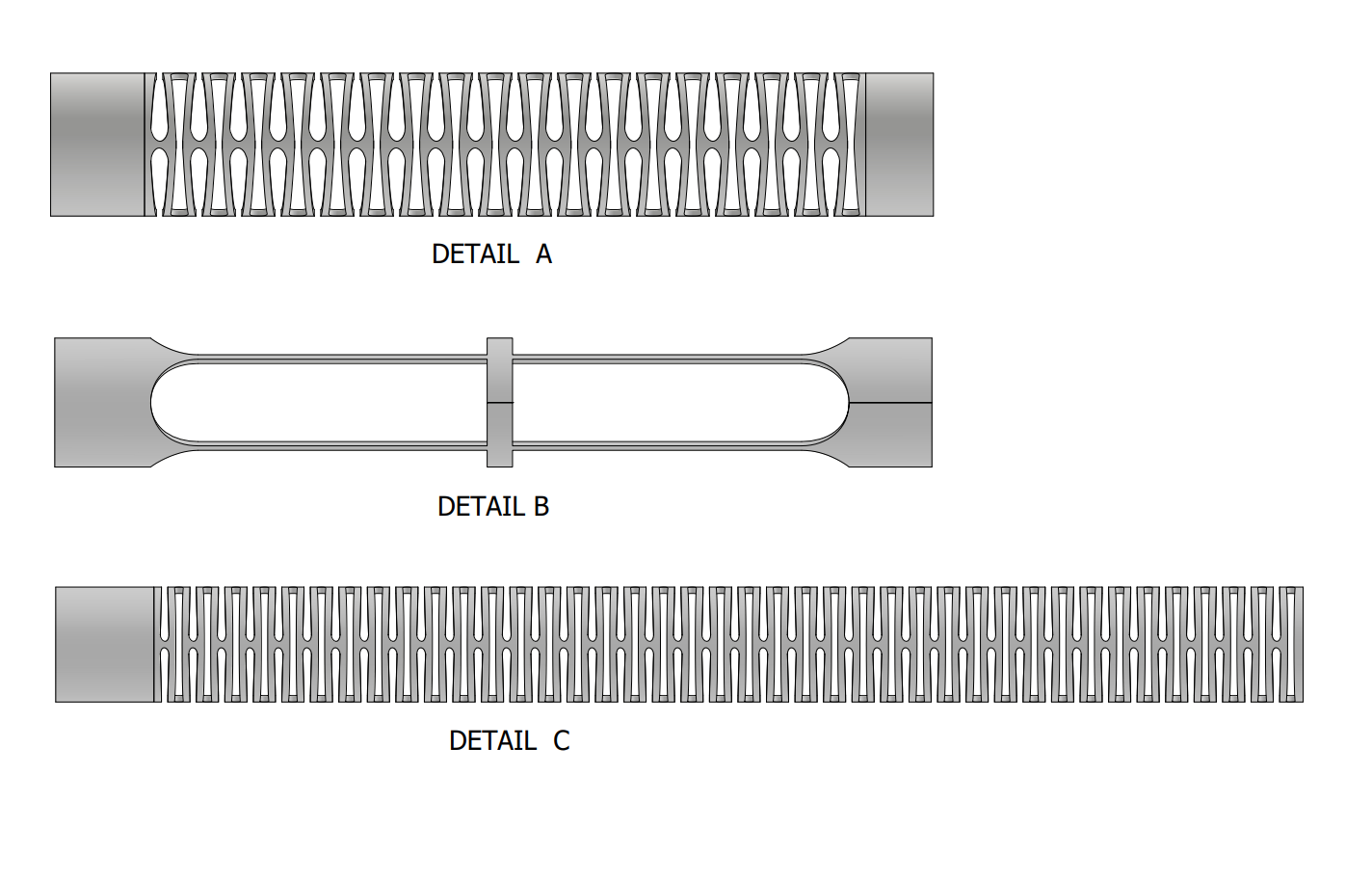"}
        \caption{Detailed section views of the nested assembly highlighting the laser lattice cutouts (Details A, B, and C).}
        \label{fig:flex_section_views}
    \end{subfigure}

    \vspace{8pt}
    \caption{The FLEX concentric tube assembly architecture: (a) exploded view of the Nitinol components and (b) sectional details of the geometrically altered laser-cut profiles.}
    \label{fig:combined_tube_assembly}
\end{figure}

The core structural body of the positioner is composed of three nested Nitinol (nickel-titanium) tubes arranged in a concentric, cascading assembly—comprising the outer tube (OT), middle tube (MT), and inner tube (IT)—and is driven by three piezo motors to achieve integrated tip, tilt, and piston actuation. This design leverages the superelastic properties of Nitinol to withstand large deflections without permanent deformation, ensuring the fiber tip remains inherently parallel with its base to preserve telecentricity across the full range of motion. As illustrated in Figure~\ref{fig:combined_tube_assembly}, each tube features a unique, laser-cut geometric flexure pattern optimized via Finite Element Analysis (FEA) to minimize localized stress concentrations and mitigate structural buckling or parasitic moments. Furthermore, the hollow internal volume allows an optical fiber to run freely along the longitudinal axis, avoiding tight bending radii during actuation. This configuration minimizes torsional stress and protects against mechanical fatigue, effectively reducing the risk of signal loss from fiber micro-fractures even when refocusing at the extreme edges of the patrol area.

\FloatBarrier

To withstand repeated large deflections without incurring permanent plastic deformation, the tube geometry was designed to utilize the superelastic properties of a binary NiTi alloy (nominally 50.8~at.\%~Ni). The alloy is specifically processed to exhibit stable superelastic behavior at the WST focal surface operating temperature via an austenite finish temperature safely below the environmental minimum ($A_f < T_{\mathrm{min}}$). Within this superelastic phase, the material can accommodate up to 8\% recoverable strain via a reversible, stress-induced martensitic transformation. To maximize the operational fatigue life of the tubes, the laser-cut lattice geometries are optimized to keep peak von~Mises stresses strictly below the lower bound of this transformation plateau ($\sigma_{\mathrm{SIM}} \approx 400$\,MPa at 22$^{\circ}$C). This constraint confines nominal operation entirely to the linear-elastic austenitic loading branch.

\begin{figure} [ht]
   \centering
   \includegraphics[width=0.9\textwidth]{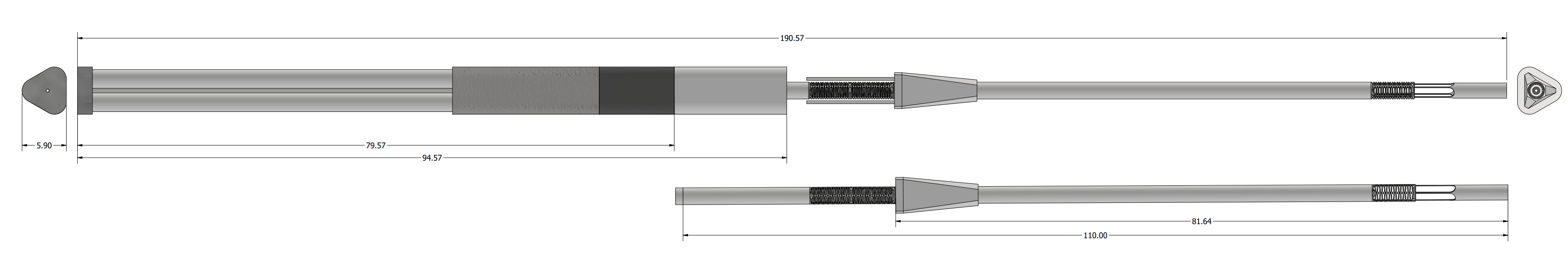}
   \caption{\label{fig:flex_tube_assembly} Complete FLEX positioner assembly, illustrating the mechanical coupling between the structural tubes and the piezo drive units.}
\end{figure}

Figure~\ref{fig:flex_tube_assembly} shows the conceptual design of the fully assembled FLEX positioner, illustrating the integration of the concentric tubes and micro-actuators. The system is $\sim$190~mm in length, and driven via three piezoelectric actuators arranged in a push-pull configuration. This setup mimics a tripod-style kinematic mount to enable three degrees of freedom (DOF), allowing the system to achieve controlled translation across all axes. This kinematic control ensures that the fiber tip can be precisely positioned anywhere within its patrol field while actively maintaining optical focus. Nitinol wires serve as the coupling links between the actuating motors and the main positioner body. During actuation, these wires conform smoothly to the bend radii of the lower tube lattice, which simultaneously reduces both the force required for displacement and stress concentrations within the lower geometric cutouts of the tubes.

\FloatBarrier

\subsection{Mechanics - Positioner Actuator}
The actuator developed for the FLEX positioner is a stick-slip piezoelectric actuator\cite{barrow2026exploration} wherein the shaft and preload have been developed with a concentric construction to maintain the tight footprint required to achieve the high packing density of positioners within the focal plane. Due to the stick-slip nature of the actuator, the drive signal of the piezoelectric takes the form of a sawtooth in order to gradually change the voltage during the stick phase, and a rapid drop in voltage during the slip phase.

Through the construction with a minimized footprint, the three actuators required for driving the FLEX positioner can be orientated in a single-plane. A stick-slip actuator offers the benefit of a no-signal hold condition, which when paired with a low voltage operation allows for a reduced power consumption per positioner relative to the comparative tilting spine positioners which operate with a much larger signal voltage. This consideration was made due to the scale of multiplexing WST is targeting hence minimizing electronic power is key. Additionally, the overall arrangement of the three actuators also reduces the complexity as it eliminates the requirement of angular movement.

\begin{figure}[ht]
    \centering

    \begin{subfigure}[b]{0.9\textwidth}
        \centering
        \includegraphics[width=\textwidth]{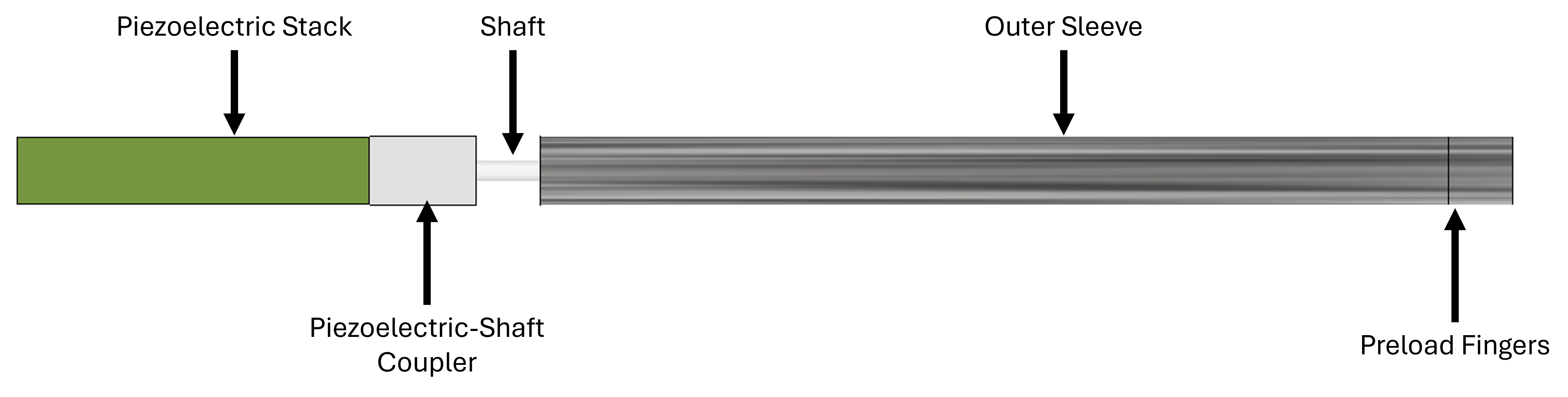}
        \caption{Side profile of the FLEX actuator assembly}
        \label{fig:actuator_full}
    \end{subfigure}
    \hfill 

    \begin{subfigure}[b]{0.9\textwidth}
        \centering
        \includegraphics[width=\textwidth]{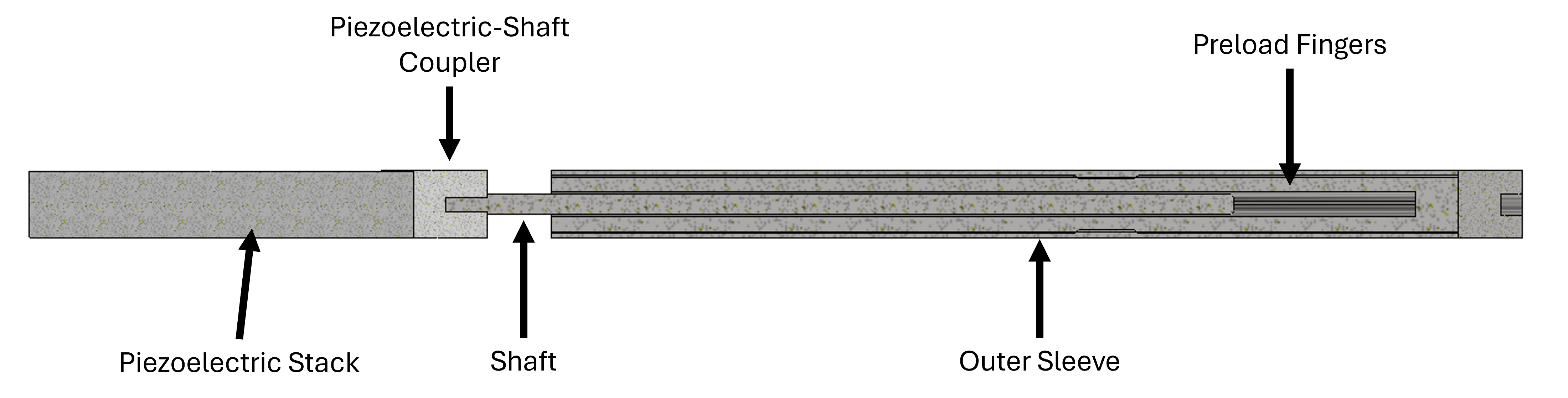}
        \caption{Side cross-section of the actuator showing preload flexure}
        \label{fig:actuator_section}
    \end{subfigure}
    
    \vspace{8pt}
    \caption{The FLEX positioner actuator design: (a) side profile assembly and (b) internal cross-section displaying the preload flexure mechanism}
    \label{fig:actuator_views}
\end{figure}

Mechanically the actuator is configured with radially deflecting fingers applying a preload force to the drive shaft, with the actuator shown in Figure \ref{fig:actuator_views}. This preload force $F_{p}$ is a function of both coefficient of friction ($\mu$) between the fingers and the drive shaft, and the required target drive force $F_{D}$. The required drive force is designed to be 1.5 times the FLEX positioners true maximum actuation force in order to build in a safety factor at the design stage. The actuators have been designed with $\mu=0.1$ to allow for a fine resolution stick-slip actuation whilst also requiring no lubrication and a fixed, non-tunable preload creating a self contained actuator.

\FloatBarrier

Using these actuators, micro positioning of FLEX with a high accuracy can be obtained by a precise control of the frequency and voltage of the waveform. The overall positioning of the FLEX positioner is based on the displacement of these actuators, where moving one or a combination of all can produce motion in FLEX in any desired direction. The control strategy is divided into two stages:

\begin{itemize}
    \item \textbf{Coarse Positioning:} The actuators operate at the maximum available voltage (24~V), producing the largest possible step size. This enables rapid movement toward the target position.
    \item \textbf{Fine Positioning:} Once the system nears the desired position, smaller adjustments are made with reduced voltage and/or controlled frequency to achieve micron-level precision and accurate final alignment
\end{itemize}

The critical physical and operational parameters of the actuator are compiled in Table \ref{tab:actuator_requirements}.

\begin{table}[ht]
    \centering
    \caption{FLEX actuator design parameters and operational constraints.} 
    \label{tab:actuator_requirements}
    \begin{tabular}{ll} 
        \toprule
        \textbf{Parameter} & \textbf{Requirement / Value} \\ 
        \midrule
        Minimum Step Size & 0.3 $\mu$m \\
        Maximum Force Output & 2.5 N  \\
        Operating Voltage Range & 0 -- 24 V \\
        Actuator Envelope & $\varnothing$ 3.2 mm $\times$ 67 mm Length \\
        Frequency Floor & $>$ 570 Hz (step size dependent) \\
        \bottomrule
    \end{tabular}
\end{table}
\FloatBarrier

\section{Simulated Positioner Performance}
\subsection{Finite Element Analysis Framework}
\begin{figure}[ht]
    \centering
    \includegraphics[width=0.9\textwidth]{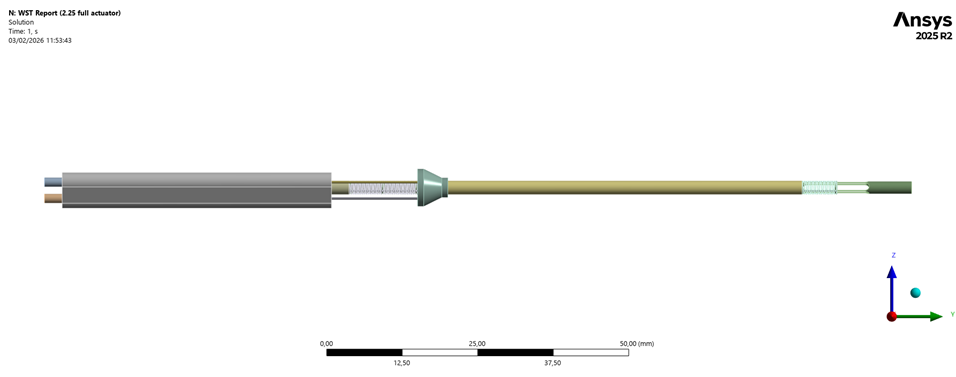}
    \caption{FEA model of the undeformed FLEX actuator assembly (2.25~mm OD) configured for ANSYS structural analysis}
    \label{fig:undeformed FEA model}
\end{figure}

To simulate the performance of the FLEX positioner, the FEA model (Figure~\ref{fig:undeformed FEA model}) was analyzed in ANSYS Mechanical. This model was used to evaluate stress concentrations within the tubes and characterize the frictional contacts between the tubes and cutout segments; these data were critical for determining the actuation forces required to reach the positioner's maximum patrol radius. Furthermore, the total displacement of the fiber tip was quantified to establish the patrol area and the resulting maximum telecentric tilt error at the stroke limits.

\FloatBarrier

The structural analysis of the FLEX positioner involved significant computational complexity due to the high density of contact surfaces, frictional nonlinearities, and the constitutive behavior of the Nitinol components. To ensure numerical stability and consistent convergence, several solver parameters and contact formulations were iteratively optimized. The specific mesh modeling and contact settings utilized in this study are detailed in Tables~\ref{tab:mesh_parameters} \&~\ref{tab:contact_parameters} respectively.

\begin{table}[ht]
    \caption{FEA Mesh Configuration Parameters} 
    \label{tab:mesh_parameters}
    \begin{center}       
    \begin{tabular}{|l|l|p{6cm}|} 
    \hline
    \rule[-1ex]{0pt}{3.5ex} \textbf{Parameter} & \textbf{Setting} & \textbf{Reasoning} \\
    \hline
    \rule[-1ex]{0pt}{3.5ex} \textbf{Global Settings} & & \\
    \hline
    \textbf{Default} & & \\
    \mc{\citem{Physics Preference}} & Nonlinear Mechanical & Accounts for Nitinol’s superelasticity, large deformations, and frictional contacts.\\
    \mc{\citem{Element Size}} & 2~mm & Provides lightweight computational baseline for rigid assembly components.\\
    \hline
    \textbf{Sizing} & & \\
    \mc{\citem{Growth Rate}} & 1.5 & Controls transition speed; balances accuracy and node count.\\
    \mc{\citem{Max Element Size}} & 4~mm & Prevents mesh from becoming too sparse in non-critical regions. \\
    \mc{\citem{Defeature Size}} & 0.025~mm & Threshold for geometric simplification to ensure mesh quality and eliminate non-convergent local singularities. \\
    \mc{\citem{Capture Curvature}} & Yes & Generates better elements for circular profiles. \\
    \mc{\citem{Curv. Normal Angle}} & 20$^{\circ}$ & Ensures smooth geometric resolution for circular profiles.\\
    \hline
    \rule[-1ex]{0pt}{3.5ex} \textbf{Thin Walled Tubes/ Features} & & \\
    \hline
    \textbf{Sweep Method} & & \\
    \mc{\citem{SRC/TRG selection}} & Automatic Thin & Facilitates hex-dominant structured meshing. \\
    \mc{\citem{Sweep Num Divis}} & 2 & Resolves stress gradients through the tube wall thickness. \\
    \hline
    \textbf{Body Sizing} & & \\
    \mc{\citem{Capture Curvature}} & Yes & Overrides global settings for high-fidelity geometry. \\
    \mc{\citem{Curv. Normal Angle}} & 12.5$^{\circ}$ -- 40$^{\circ}$ & Refines bending zones while optimizing straight segments.\\
    \hline
    \textbf{Face Sizing} & & \\
    \mc{\citem{Element Size}} & 0.05 -- 0.15~mm & Required to resolve high-pressure contact patches. \\
    \mc{\citem{Face Meshing}} & Tube faces & Ensures uniform node patterns for stable frictional sliding and contact bonds between tubes. \\
    \hline
    \rule[-1ex]{0pt}{3.5ex} \textbf{Resulting Mesh} & & \textbf{Quality Assessment}\\
    \mc{\citem{Total Nodes}} & 1,417,417 & Resolution required for convergence of nonlinear Nitinol behavior. \\
    \mc{\citem{Total Elements}} & 316,016 & Discretization sufficient for resolving assembly-level clearances. \\
    \mc{\citem{Avg. Element Quality}} & $\sim$0.8 & Indicates high geometric regularity and numerical stability. \\
    \mc{\citem{Skewness Average}} & $\sim$0.19 & Confirms minimization of discretization errors. \\
    \mc{\citem{Skewness Std. Dev.}} & $\sim$0.16 & Confirms high uniformity and absence of outlier elements.\\
    \hline
    
    \end{tabular}
    \end{center}
\end{table}
\FloatBarrier

\begin{table}[ht]
    \caption{FEA Frictional Contact Configuration Parameters} 
    \label{tab:contact_parameters}
    \begin{center}       
    \begin{tabular}{|l|l|p{6cm}|} 
    \hline
    \rule[-1ex]{0pt}{3.5ex} \textbf{Parameter} & \textbf{Setting} & \textbf{Reasoning} \\
    \hline
    \rule[-1ex]{0pt}{3.5ex} \textbf{Frictional Contact Settings} & & \\
    \hline
    \mc{\citem{Frictional Coefficient}} & 0.1 & Simplifies convergence by minimizing stick-slip oscillations at sliding interfaces.\\
    \mc{\citem{Trim Contact}} & Off & Prevents the solver from ignoring contacts that become active during large-scale displacements.\\
    \mc{\citem{Formulation}} & Pure Penalty  & Enhances numerical robustness for large-sliding nonlinear contact problems.\\
    \mc{\citem{Detection Method}} & On Gauss Point & Ensures smooth pressure distribution across curved cylindrical surfaces.\\
    \mc{\citem{Penetration Tolerance Value - Value}} & 0.005~mm & Balances geometric accuracy with solver stability for thin-walled components.\\
    \mc{\citem{Elastic Slip Tolerance - Value}} & 0.03~mm & Provides numerical damping to stabilize transitions between stuck and sliding states.\\
    \mc{\citem{Normal Stiffness Factor}} & 0.1 & Reduces contact "hardness" to facilitate convergence in highly nonlinear models.\\
    \mc{\citem{Update Stiffness}} & Each iteration, Aggressive & Accounts for the shifting stiffness of Nitinol during phase transformations.\\
    \mc{\citem{Stabilization Damping Factor}} & 0.01 & Suppresses initial rigid-body motion before full contact is established.\\
    \mc{\citem{Pinball Region - Radius}} & 0.05~mm & Restricts the contact search to the physical clearance between assembly parts.\\
    \hline
    
    \end{tabular}
    \end{center}
\end{table}
\FloatBarrier

\begin{figure}[ht]
    \centering
    \includegraphics[width=0.9\textwidth]{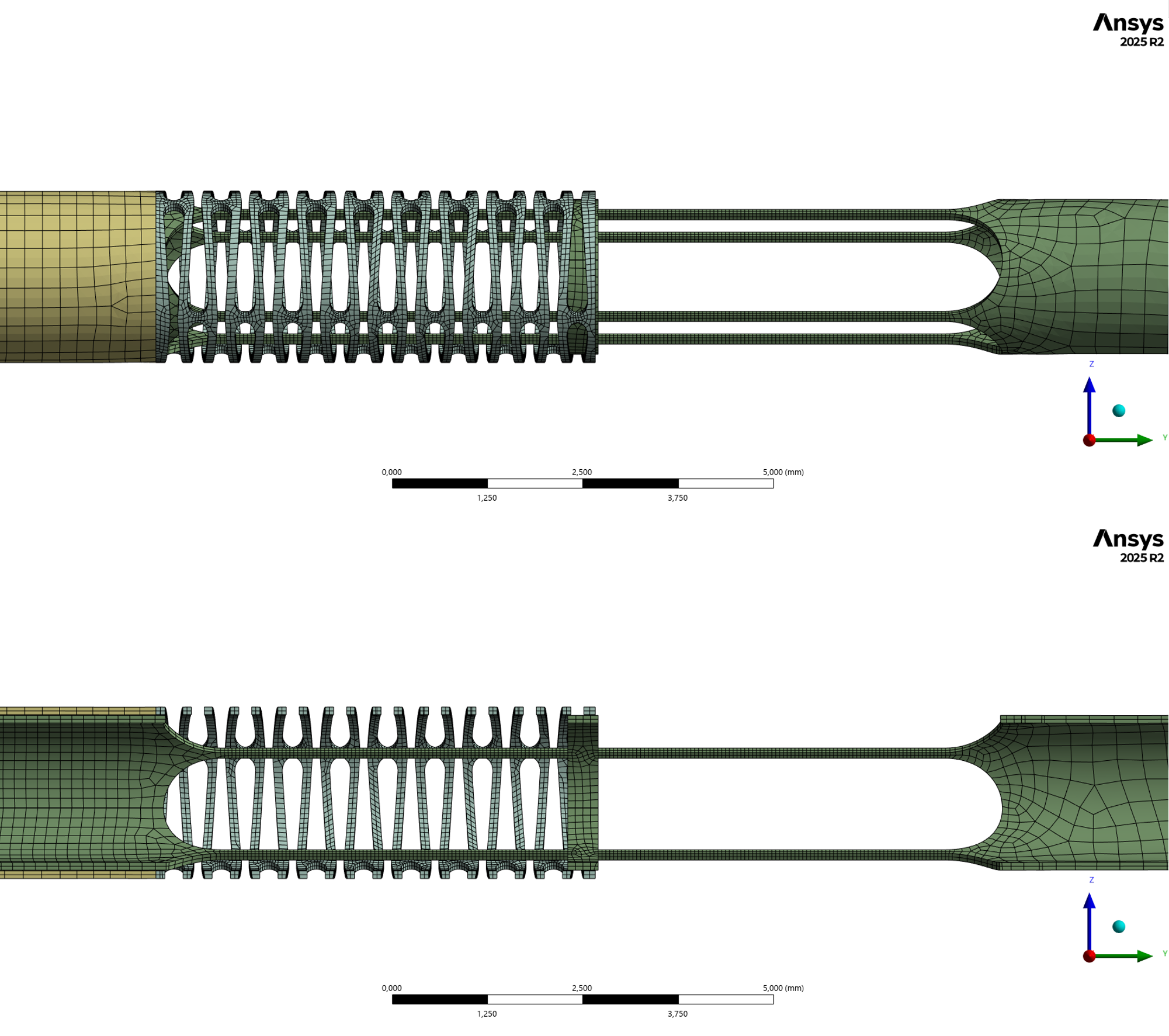}
    \caption{Detail view of the FEA mesh for the positioner upper lattice section, illustrating the surface discretization (top) and the internal through-thickness mesh via cross-section (bottom).}
    \label{fig:FEA mesh detail view}
\end{figure}

These settings yield a high-fidelity mesh with a minimum of two elements across the thinnest features to accurately resolve stress concentrations. Furthermore, the chosen contact configurations allow for precise detection of frictional interfaces, which is essential for determining the total required actuation force without impacting convergence. Detail views of the mesh are presented in Figure~\ref{fig:FEA mesh detail view}.

\FloatBarrier

\subsection{Results and Discussion}
\begin{figure}[ht]
    \centering
    \includegraphics[width=0.9\linewidth]{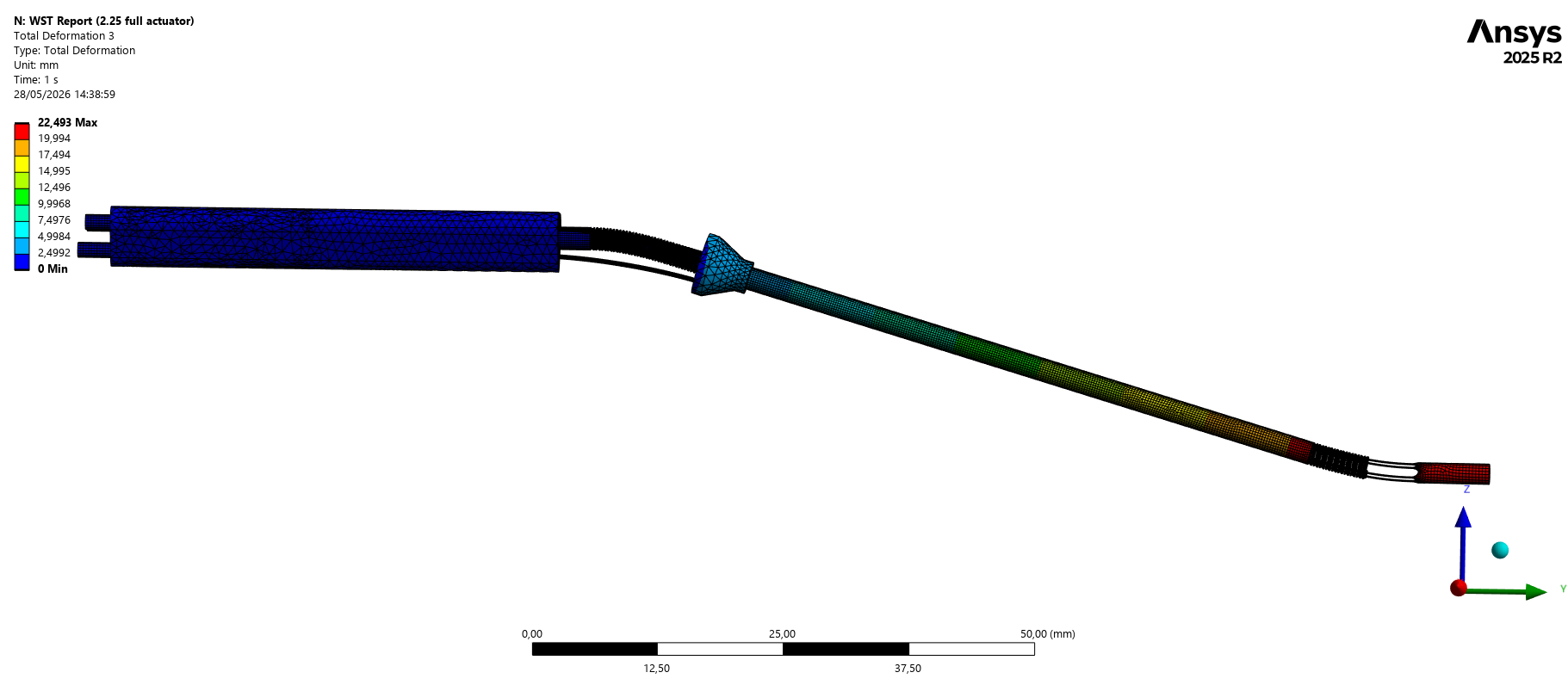}
    \caption{Simulated structural deformation of the FLEX positioner under a differential 0.4~mm axial stroke (Motor 1: -0.4mm down; Motor 2 \& 3: +0.4 mm), where maximum deformation occurs at the fiber tip ($\sim$22.5~mm displacement)}
    \label{fig:simulated structural deformation}
\end{figure}

The model was then simulated when actuated 0.4~mm in a push-pull axial differential configuration (with two Nitinol wires that are in-plane with each other pushing in the same direction, and the third wire pulling in the opposite direction). The resulting deformation can be seen in Figure~\ref{fig:simulated structural deformation}, where the maximum displacement was found to be $\sim$22.5mm, when operating below the Nitinol stress transformation plateau of $\sim$400 MPa. At the maximum displacement, the telecentric tilt angle seen at the fiber tip was $\sim$0.385$^{\circ}$. 

\FloatBarrier

\subsubsection{Stress and Fatigue}

\begin{figure}[ht]
    \centering
    \includegraphics[width=0.9\linewidth]{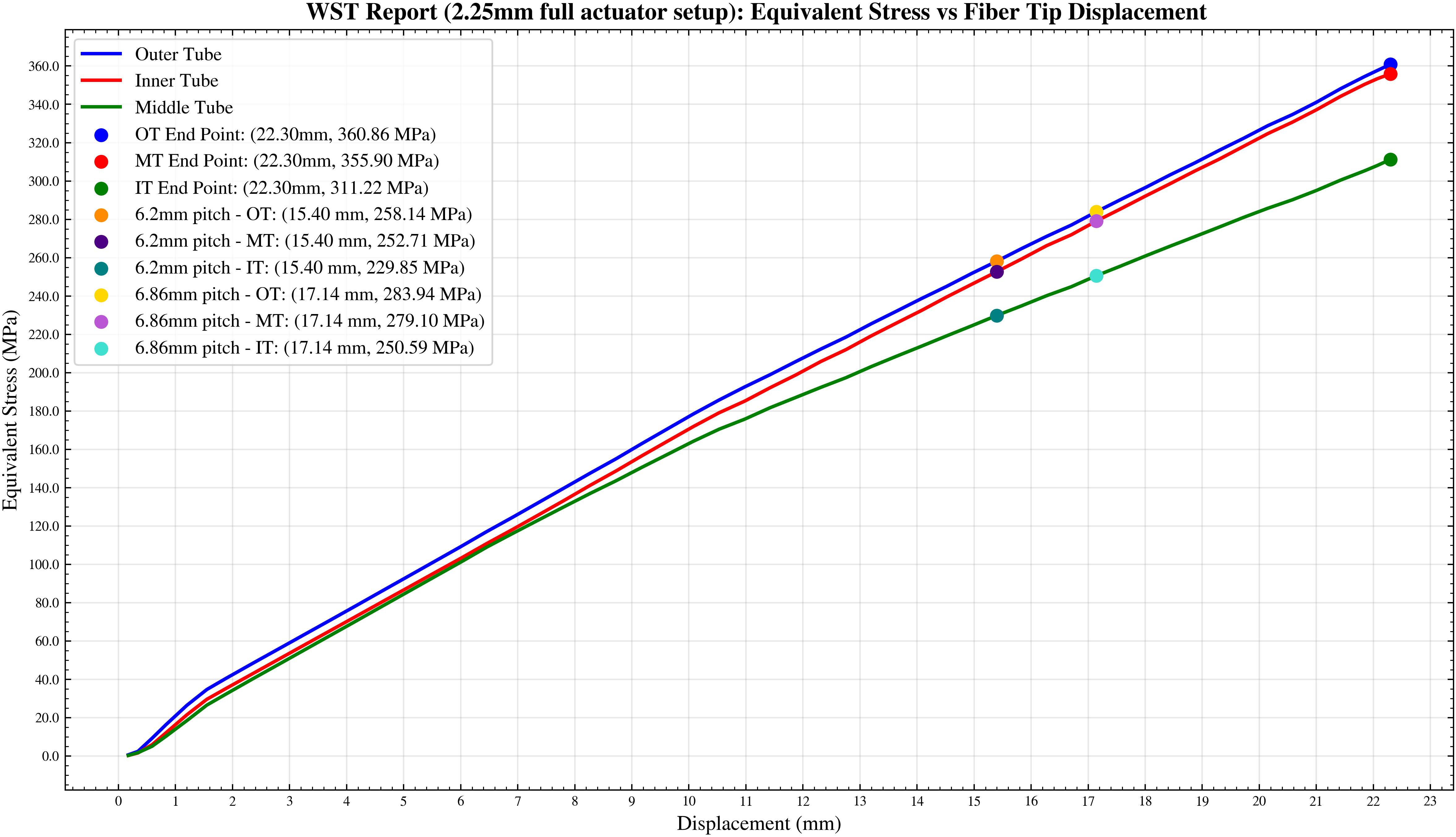}
    \caption{Simulated peak von~Mises equivalent stress in the OT, MT, and IT as a function of fiber tip displacement (patrol radius). All values remain below the Nitinol superelastic transformation plateau upper bound of ${\sim}400$~MPa, confirming operation within the recoverable strain regime.}
    \label{fig:Equivalent Stress vs Displacement}
\end{figure}

Figure~\ref{fig:Equivalent Stress vs Displacement} shows the peak von~Mises equivalent stress in each Nitinol tube as a function of fiber tip displacement. At maximum actuation, the stresses in the outer tube~(OT), middle tube~(MT), and inner tube~(IT) are 360.9~MPa, 355.9~MPa, and 311.2~MPa, respectively. All three values remain below the upper bound of the stress-induced martensitic transformation plateau (${\sim}400$~MPa at 22$^{\circ}$C), indicating operation within the superelastic loading branch. This avoids plastic deformation and is consistent with recoverable actuation behavior, which is a prerequisite for cyclic durability. The stresses for each tube at both the set 6.2~mm and 6.86~mm pitch can be seen in Table \ref{tab:stress_pitch}.

\FloatBarrier

\begin{table}[ht]
    \centering
    \caption{Peak von Mises equivalent stress in Nitinol tubes at respective 2.5x pitch patrol radius}
    \label{tab:stress_pitch}
    \begin{tabular}{@{}lcc@{}}
        \toprule
        \textbf{Tube Type} & \textbf{p = 6.2~mm} & \textbf{p = 6.86~mm} \\
        \midrule
        Outer Tube  & 258.14~MPa & 283.94~MPa \\
        Middle Tube & 252.71~MPa & 279.10~MPa \\ 
        Inner Tube  &  229.85~MPa & 250.59~MPa \\ 
        \bottomrule
    \end{tabular}
\end{table}

While classical fatigue models, such as the Coffin--Manson--Morrow relationship, predict exceptionally long fatigue lives exceeding $10^8$ cycles for these configurations, they are not well suited for superelastic NiTi. Conventional equations typically model macro-plastic slip accumulation; however, the fatigue life of Nitinol is driven by the localized stress of repeatedly shifting between the austenite and martensite crystal phases. This cyclic transformation causes micro-cracks to form around internal defects, fundamentally differing from the structural dislocation pile-ups and cyclic plastic slip typical of conventional ductile metals~\cite{Pelton2008}.

Originally established from rotating-beam fatigue testing of superelastic specimens, the Pelton framework characterizes high-cycle fatigue life using two parameters derived from the FEA-extracted peak strains: 

\begin{equation}
\text{Mean strain:} \hspace{5pt} \varepsilon_m = \frac{\varepsilon_{\max} + \varepsilon_{\min}}{2},
\end{equation}
\begin{equation}
\text{Strain amplitude:} \hspace{5pt} \varepsilon_a = \frac{\varepsilon_{\max} - \varepsilon_{\min}}{2}.
\end{equation}
Here, $\varepsilon_{\max}$ represents the peak strain at the maximum positioner patrol radius, while $\varepsilon_{\min}$ accounts for the home position residual assembly strain. The operational design boundary is defined by a conservative $10^7$-cycle limit of $\varepsilon_{a,\mathrm{lim}} \approx 0.40\%$ at zero mean strain, which degrades linearly as $\varepsilon_m$ increases~\cite{Pelton2008}.

\begin{figure}[ht]
    \centering
    \includegraphics[width=0.9\linewidth]{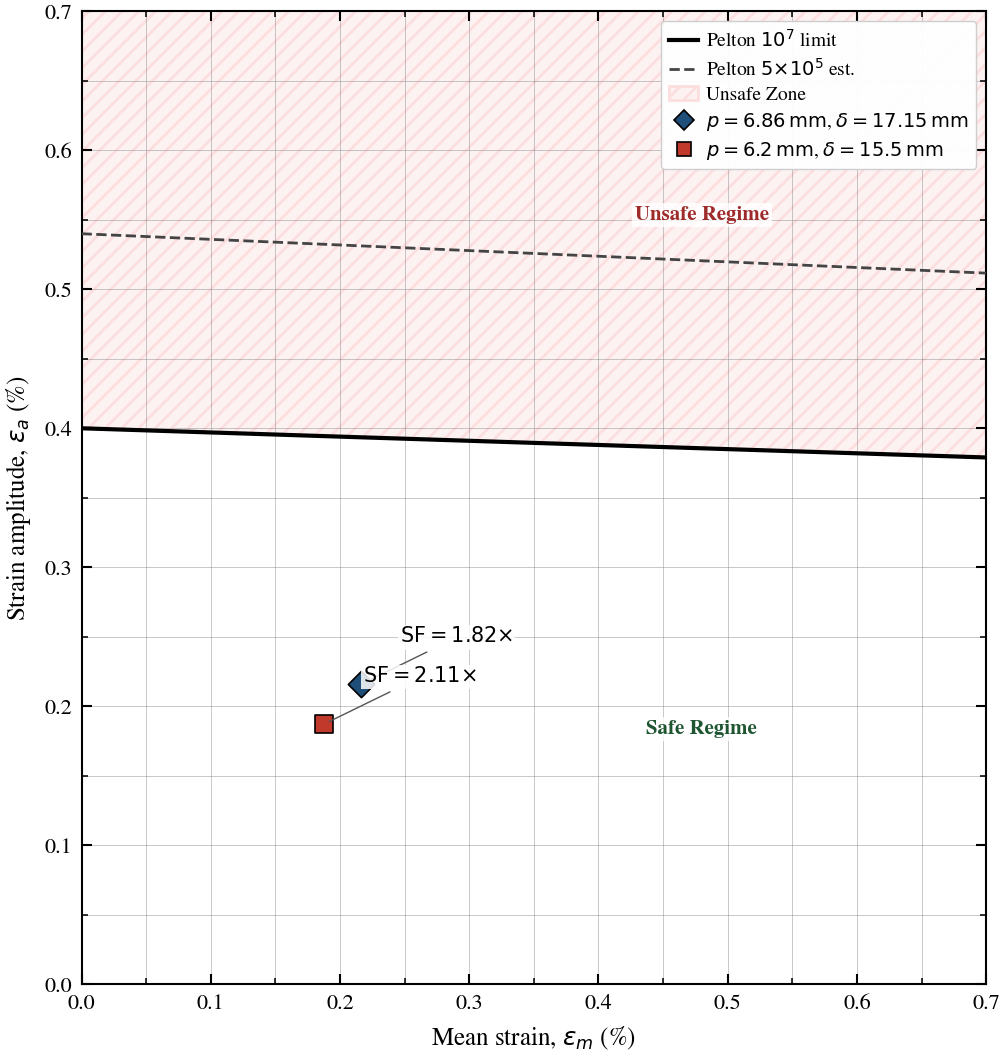}
    \caption{Pelton mean strain/strain amplitude fatigue diagram for the FLEX Nitinol tube assembly~\cite{Pelton2008}. The solid boundary is the conservative $10^7$-cycle limit for superelastic NiTi; the dashed boundary is an estimated $5 \times 10^5$-cycle limit. Design points for both pitch configurations lie within the safe regime, with fatigue margins of $1.82\times$ ($p = 6.86$~mm) and $2.11\times$ ($p = 6.2$~mm) relative to the $10^7$-cycle boundary.}
    \label{fig:pelton_fatigue}
\end{figure}

The operating strain amplitudes and mean strains derived from the FEA peak strain values for each configuration are summarized in Table~\ref{tab:pelton_fatigue}. As shown in Figure~\ref{fig:pelton_fatigue}, both design points fall within the safe regime below the $10^7$-cycle Pelton boundary, suggesting that the tube assemblies have the potential to exceed $10^7$ operational cycles without fatigue failure.

\begin{table}[ht]
    \centering
    \caption{Pelton framework fatigue assessment for the FLEX Nitinol tube assembly. Peak strains are derived from FEA at maximum patrol radius under a zero-based actuation cycle. The Fatigue Margin is defined as $\mathrm{FM} = \varepsilon_{a,\mathrm{lim}}(\varepsilon_m) / \varepsilon_a$.}
    \label{tab:pelton_fatigue}
    \begin{tabular}{@{}lcccccc@{}}
        \toprule
        \textbf{Configuration} & $\varepsilon_{\max}$ & $\varepsilon_m$ & $\varepsilon_a$ & $\varepsilon_{a,\mathrm{lim}}$ & \textbf{FM} \\
         & (\%) & (\%) & (\%) & (\%) & & \\
        \midrule
        $p = 6.86$~mm, $\delta = 17.15$~mm & 0.432 & 0.216 & 0.216 & 0.394 & $1.82\times$ \\
        $p = 6.2$~mm, $\delta = 15.5$~mm & 0.375 & 0.187 & 0.187 & 0.394 & $2.11\times$ \\
        \bottomrule
    \end{tabular}
\end{table}

\FloatBarrier

This predicted fatigue life is substantially greater than the WST requirement of 500,000 reconfiguration cycles. These results represent a simulation-based assessment, where the fatigue assessment was conducted under several idealized assumptions:
 
\begin{enumerate}[label=(\alph*)]
    
    \item The Pelton limit is derived from polished wire and tube specimens, whereas the FLEX tubes are laser-cut, introducing a recast layer and heat-affected zone that reduce local fatigue strength relative to the bulk material. To minimize this effect, post-laser-cut electropolishing may be required to restore the native oxide passivation layer before the calculated margins can be considered verified~\cite{Pelton2008}.
    
    \item The analysis was conducted at a nominal operating temperature of 22$^{\circ}$C, and must be repeated across the full WST focal surface thermal envelope to confirm that the austenite finish temperature $A_f$ remains below the minimum operational temperature to ensure stable superelastic behavior throughout the mission.
\end{enumerate}

\subsubsection{Translational and Rotational Deviations:}
\begin{figure}[ht]
    \centering
    \includegraphics[width=0.9\linewidth]{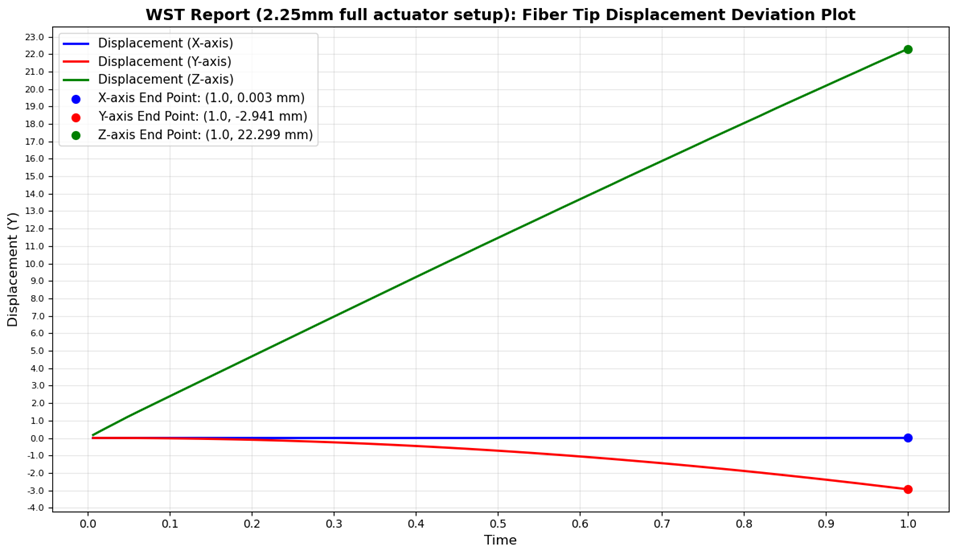}
    \caption{Plot of fiber tip displacement (patrol radius) as a function of time (normalized) in respective XYZ axes, where time depicts simulation steps towards a converged solution}
    \label{fig:displacement deviation}
\end{figure}

To evaluate the positioner’s orthogonality relative to the focal plane, both its displacement and rotation deviations were analyzed to determine if there was significant parasitic motion during actuation. 
As seen in Figure \ref{fig:displacement deviation}, as the positioner moves along the intended Z-axis, there is little movement or deviation in the X-direction; confirming high linear fidelity with minimal lateral wobbling. The decrease in the Y-axis represents the defocus in the system as the positioner tilts. This defocus is corrected by driving all three actuators in the same axial direction to realign the fiber tip with the focal plane.

\FloatBarrier

\begin{figure}[ht]
    \centering
    \includegraphics[width=0.9\linewidth]{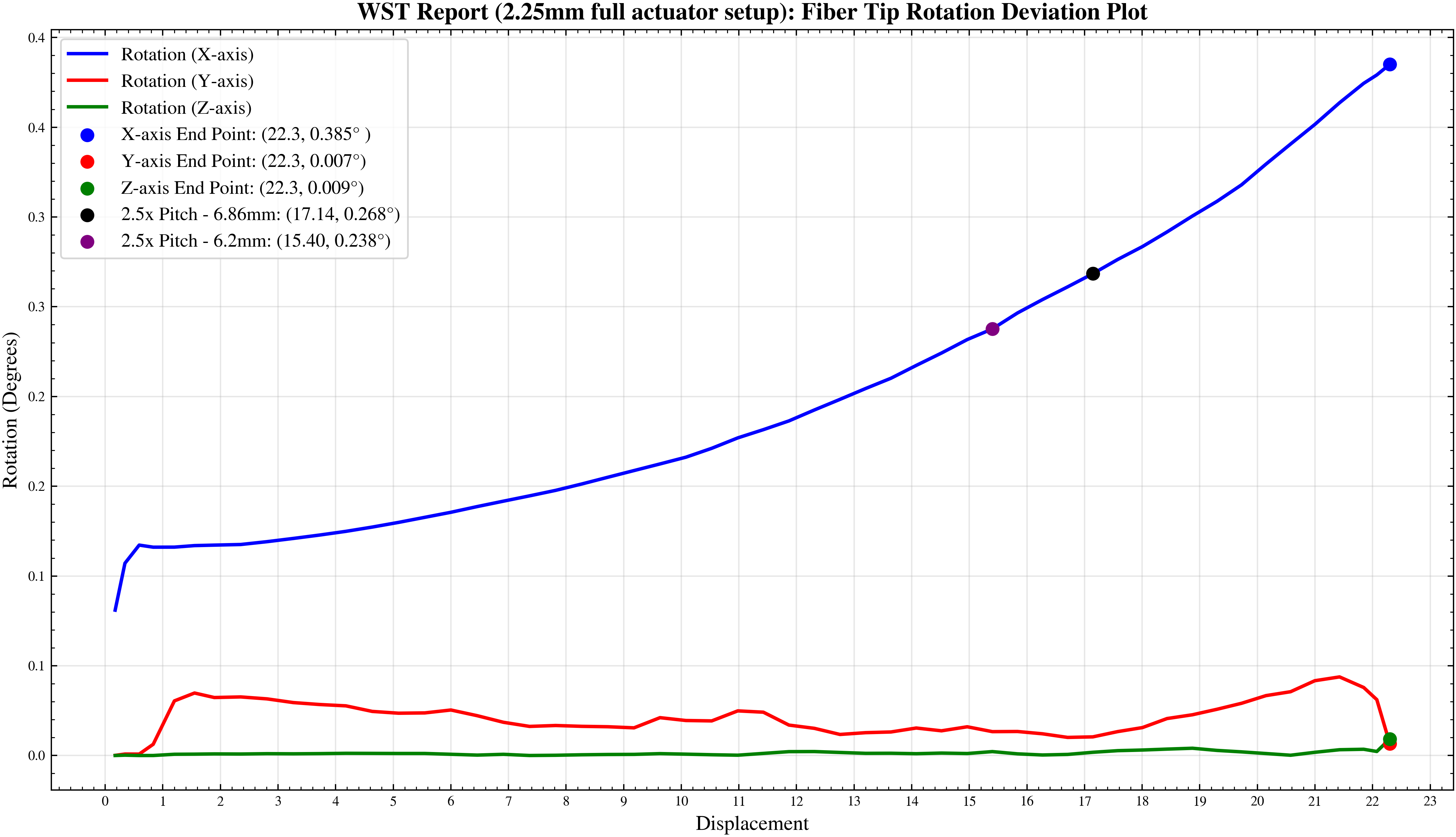}
    \caption{Plot of fiber tip rotation as a function of displacement (patrol radius) in respective XYZ axes}
    \label{fig:rotation deviation}
\end{figure}

The rotation deviation study, seen in Figure \ref{fig:rotation deviation}, shows that as the positioner moves along the intended Z-axis, the largest rotation occurs about the X-axis. This measurement corresponds to the positioner’s tilt when deviating from orthogonal with respect to the focal plane. This behavior is expected, and the positioner tubes were designed to minimize this effect. As such, at the requirement for 2.5x pitch ($\sim$15.5~mm \& $\sim$17.15~mm), the positioner sees a rotation of $\sim$0.238° and $\sim$0.268° respectively, both under the defined maximum of 0.5°. There is some torsional rotation about the Y-axis (peaking at $\sim$0.044°), likely due to frictional interference between tubes as the positioner deforms, but this phenomenon is relatively stable throughout the actuation cycle.

\FloatBarrier

\subsubsection{Actuation Force Requirement:}
The mechanical performance of the FLEX assembly was evaluated to ensure compatibility with the available actuation overhead. Given the current configuration, the actuation force required to reach the target patrol radius of $\sim$15.5~mm and $\sim$17.15~mm was approximately 1.17~N and 1.32~N respectively. This total accounts for the internal strain of the Nitinol tubes as well as parasitic losses, such as friction within the concentric interfaces and the bottom mounting assembly. To maintain a factor of safety (FoS) of $\sim$2 at maximum extension relative to these preliminary results, the actuator driving force has been specified at 2.5~N.

\begin{figure}[ht]
    \centering
    \includegraphics[width=0.9\linewidth]{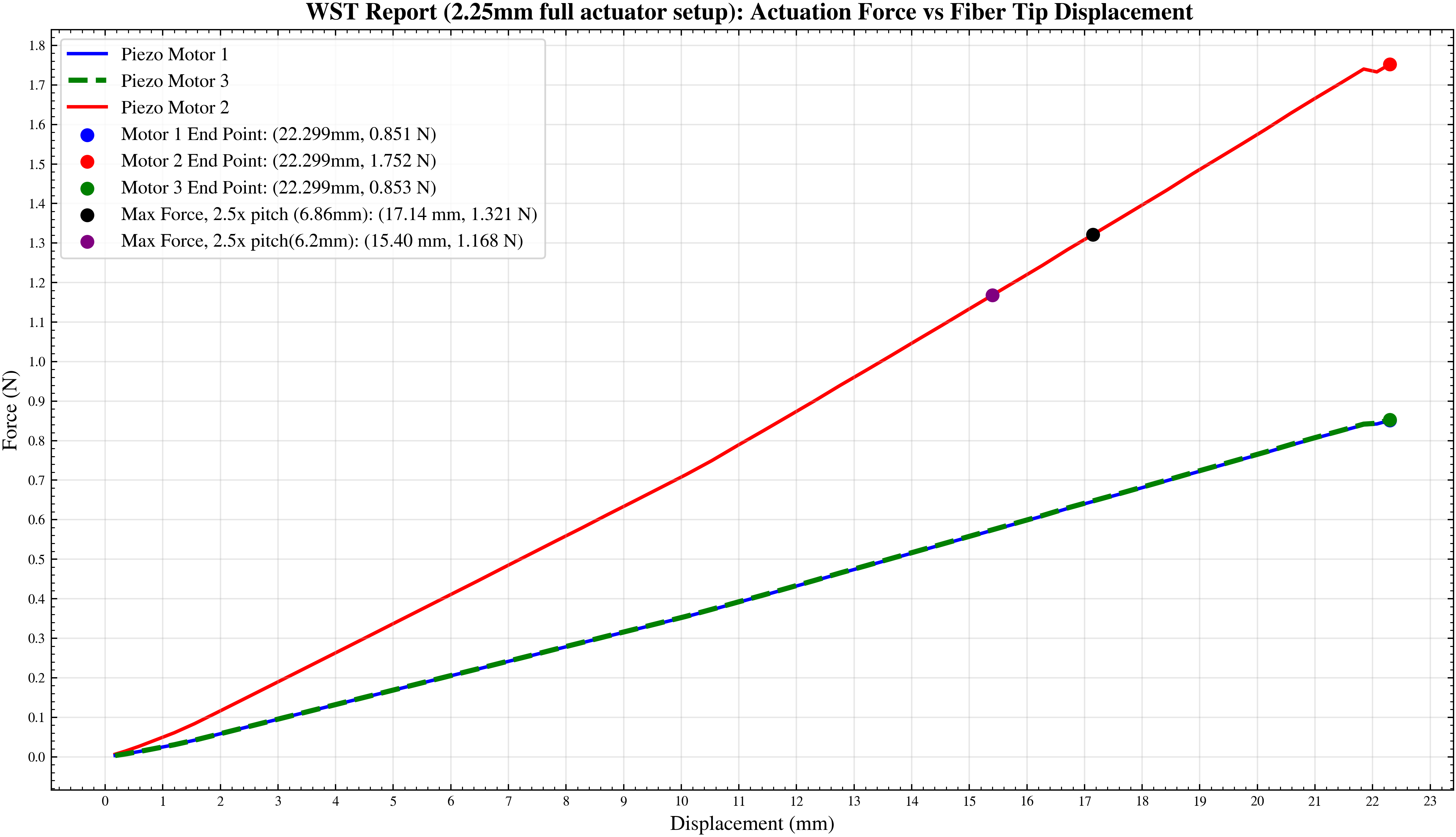}
    \caption{Plot of actuation force profile as a function of positioner tip displacement}
    \label{fig:actuation force}
\end{figure}
\FloatBarrier

\subsection{Summary and Future Work}

Table \ref{tab:performance summary} provides an overview of the simulated performance of the FLEX positioner. Based on the design criteria listed in Table \ref{tab:technical requirements}, all simulated parameters successfully meet or exceed the technical specifications. 

\begin{table}[ht]
    \centering
    \caption{FLEX positioner simulated performance summary} 
    \label{tab:performance summary}
    \begin{tabular}{@{}lcc@{}} 
        \toprule
        \textbf{Parameter} & \textbf{@ 6.86~mm pitch} & \textbf{@ 6.2~mm pitch} \\ 
        \midrule
        Patrol Radius & $\sim$17.15~mm & $\sim$15.5~mm \\
        Telecentricity Error & $\approx 0.268^{\circ}$ & $\approx 0.238^{\circ}$ \\
        Maximum Equivalent Stress & 283.9~MPa & 258.1~MPa \\
        Actuation Force & 1.32~N & 1.17~N \\
        Lifetime Reconfigurations & $> 500,000$ cycles & $> 500,000$ \\
        Positioning Accuracy\textsuperscript{*} & TBD & TBD \\
        \bottomrule      
        \multicolumn{3}{l}{\footnotesize \textsuperscript{*} Positioning accuracy to be determined in future experimental work.} \\
    \end{tabular}
\end{table}

Compared to the established architectures in Table~\ref{tab:positioner_architectures_comparison}, the FLEX system delivers a clear performance advantage. It maintains a maximum telecentric error that is competitive with high-precision theta-phi systems, but is able to leverage its large patrol radius and active focusing capabilities to eliminate tilt-induced throughput degradation. When deployed at the tight mounting pitches demanded by high-density packing, these overlapping patrol fields provide exceptional geometric redundancy. This allows for tight target clustering and maximizes the final fiber allocation fraction in highly crowded fields.

Future work will focus on evaluating the positioning accuracy of the FLEX mechanism to validate compliance with the requirement of a $\leq$15~$\mu$m positioning error for $\geq$98\% of all active fibers. Additionally, experimental studies will be conducted using physical prototypes to validate the structural behaviors observed in finite element simulations and to characterize any unmodeled parasitic movements, such as stick-slip phenomena or micro-frictional variations occurring between the sliding tube interfaces during actuation. Finally, the electromechanical integration between the actuators and the positioner tubes will be addressed, focusing on detailed mechanical housing designs and robust coupling implementations. These assemblies will be optimized for ease of fabrication and assembly to ensure scalability for the required production of over 30,000 positioners.

\section{Focal Surface Module Design}
\subsection{Overview}
A hexagonal positioner layout is convenient for WST, because (a) it allows tiling of the sky, (b) it leaves
space within the field of view for guide and wavefront-sensing cameras, and (c) the S/N declines rapidly
towards the outer parts of the field (image quality, chromatic distortion, vignetting, positioning errors,
differential distortion are all degrading), reducing the utility of positioners towards the edge of the field.

The Prime Focus Spectrograph (PFS) instrument~\cite{Tamura2016} pioneered the concept of identical linear modules
tiling the focal surface in a 'triskele' pattern with three identical sections or ‘trients’ (like quadrants, but
only three of them). The only loss of fill factor is due to a Y-shaped support strut. But the PFS focal
surface is flat, the modules are straight, and the fibers are far from pupil-centric. For the Sphinx
positioner for the proposed Maunakea Spectroscopic Explorer (MSE) project~\cite{Smedley2018}, a module shape was
devised which allows a curved focal surface to be tiled by identical modules, again giving a hexagonal
layout. The module edges form sections of great circles, so the modules are slightly tapered between
inner and outer ends.

\subsection{Tiling Curvature and Geometric Layout}

\begin{figure}[ht]
    \centering
    \includegraphics[width=1\linewidth]{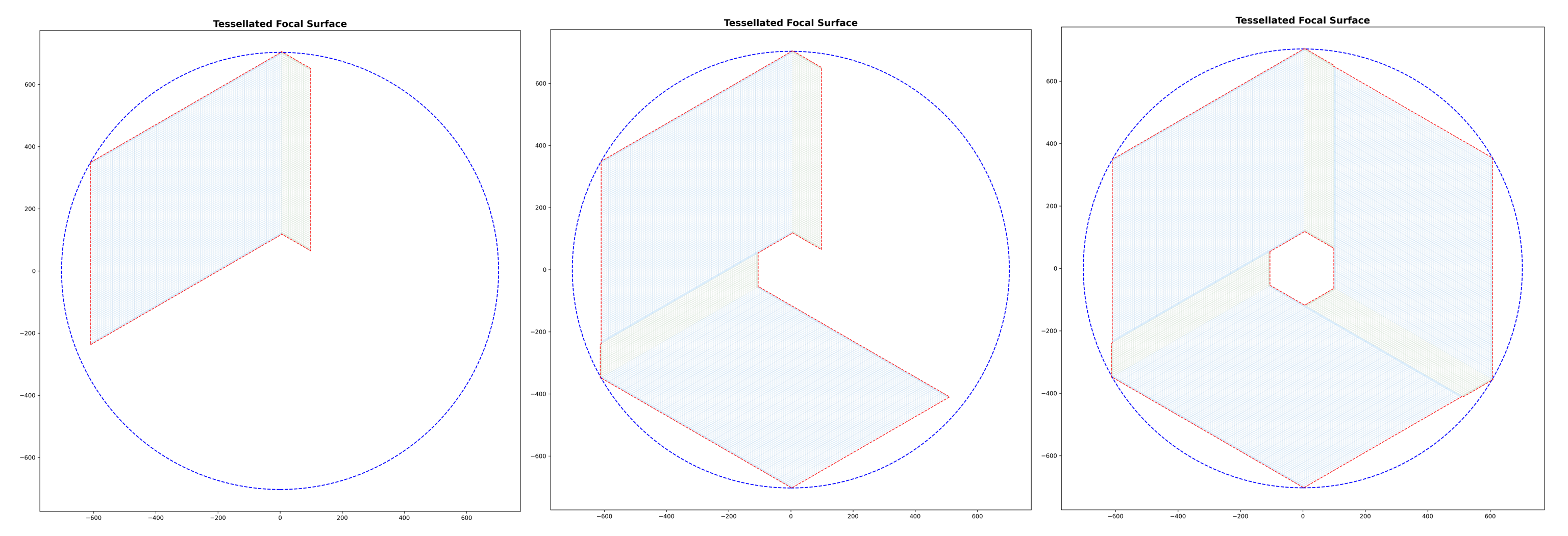}
    \caption{Trient grouping of modules (left to right: trient 1 $\rightarrow$ trient 1 + 2 $\rightarrow$ trient 1 + 2 + 3)}
    \label{fig:trient tiling}
\end{figure}

The layout for WST is further complicated by the presence of the IFS M3 pickoff mirror at the center. A modification of the triskele layout was devised in 2025, giving a hexagonal space at the field center; the modules remain identical, except that a minority are 'left-handed' (LH), mirror images of the 'right-handed' (RH) majority. The Y-struts are now offset, with an additional hexagonal support ring at the center. This layout no longer gives an exact tiling at the position of the struts, but the maximum error is only $\sim$60~$\mu$m, easily taken up in the $\sim$6~mm width of the support struts.
Each trient consists of 26 RH and 4 LH modules (Figure \ref{fig:trient tiling}). 30 modules per trient ensures symmetric
modularity for fiber routing to spectrographs.

\FloatBarrier

\begin{figure}[ht]
    \centering
    \includegraphics[width=1\linewidth]{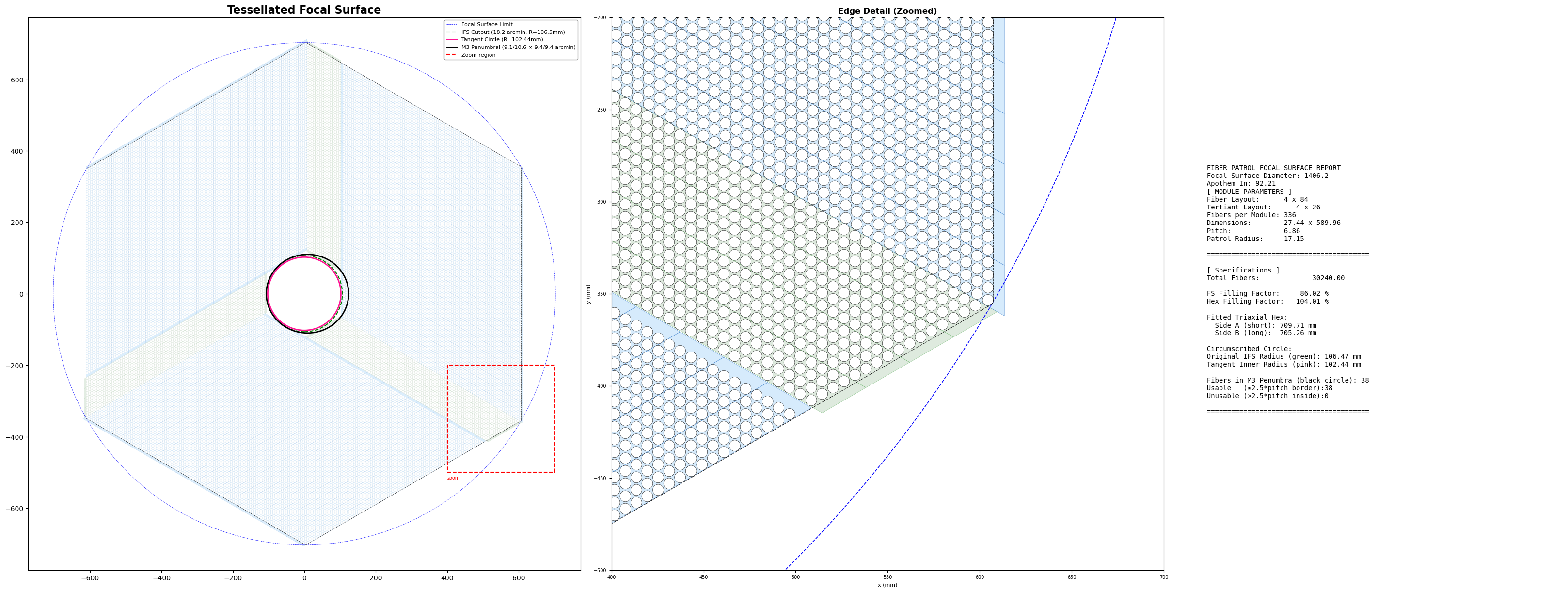}
    \caption{Top view of the fully tiled focal surface (left) and zoomed view of trient transition (middle), alongside summary
table of focal surface parameters (right)}
    \label{fig:focal surface - zoomed}
\end{figure}

When fully tiled together, the modules form the layout seen in Figure \ref{fig:focal surface - zoomed}. The geometrical constraints of module and strut width mean that inner and outer hexagonal shapes are not regular hexagons, but they are 'semi-regular' (alternating sides with 2 different lengths). The irregularity is small – 16.5 vs 18.5 pitches at the center, and 101.5 vs 103.5 at the outside. The overall semi-regular hexagonal shape means that the tiles do not quite tessellate on the sky (even when neglecting sky curvature), there are single-positioner holes between tiles.

\subsection{LR/ HR Fiber Layout}

Given that the FLEX positioner has a patrol radius of up to $\sim$2.5 x pitch, an HR fraction of 1/16 then (uniquely)
gives about the right number of HR fibers, with full field of view coverage, and with identical modules.
Each module is 4 rows wide and 84 positioners long, giving 21 HR and 315 LR positioners, with an HR
positioner at every 4th position in a single row. For the WST layout, in total, there are 30240 fibers (1890 HR + 28350 LR).

\begin{figure}[ht]
    \centering
    \includegraphics[width=0.9\linewidth]{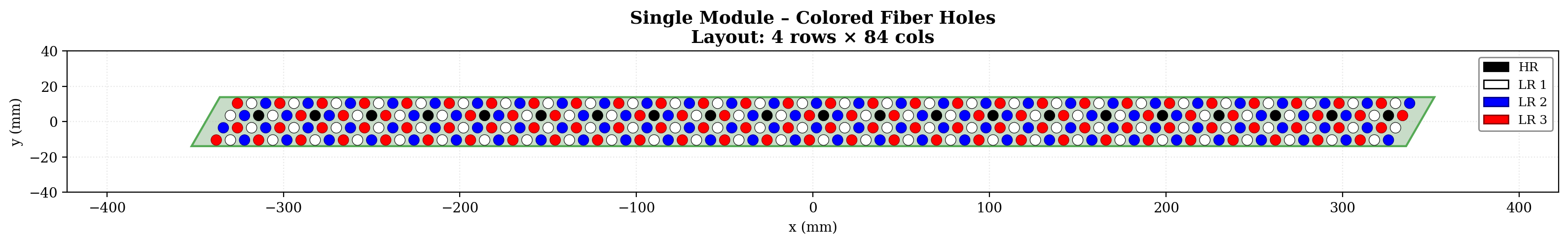}
    \caption{ Fiber distribution layout for a single LH module. The HR fiber placement is illustrative; they can be in any row and
start at any position, to allow full FoV coverage. For RH modules, the HR row would be changed so as to maintain the pattern.}
    \label{fig:module - HR LR}
\end{figure}

The LR fibers from each module are arranged into 21$\times$15-fiber slitlets of which 7 can be fed to three spectrographs. The fibers are selected as shown in Figure \ref{fig:module - HR LR}, giving full sky coverage for the fibers for each spectrograph. This means that the targets can be allocated to a spectrograph according to their magnitude, reducing the variation within each spectrograph and the resulting cross-talk issues.

\FloatBarrier

\begin{figure}[ht]
    \centering
    \includegraphics[width=0.8\linewidth]{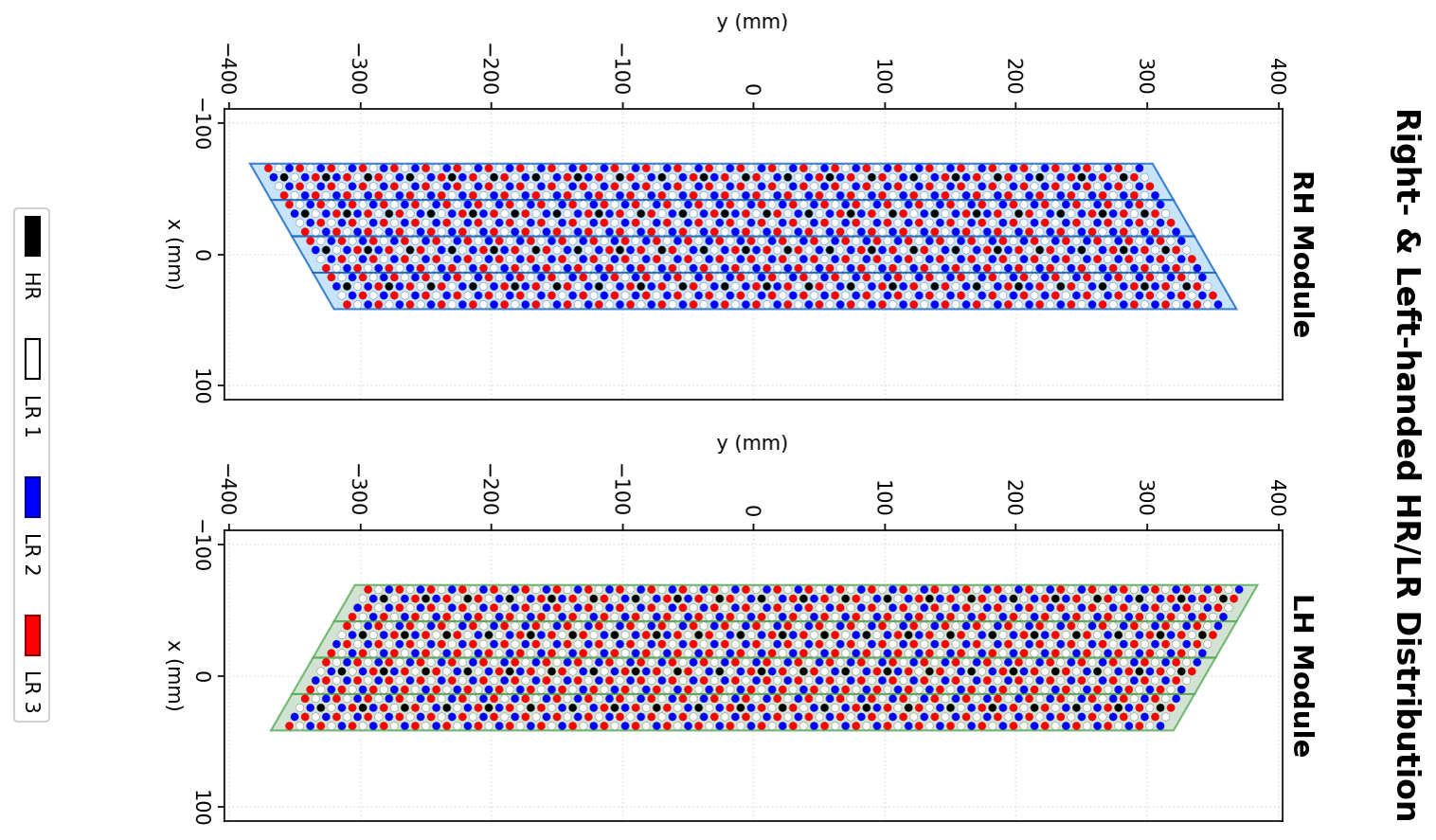}
    \caption{ Fiber distribution layout across four RH \& LH modules. Note that each LR spectrograph has an unbroken hexagonally-packed fiber layout – for example, the blue, white, red, blue fibers from the 4 modules in the lower picture, in that order from top to bottom, would be fed to the same spectrograph.}
    \label{fig:module - HR LR - multiple}
\end{figure}

Figure \ref{fig:module - HR LR - multiple} shows the HR \& LR fiber placement across tiled modules, as well as their focal surface coverage when coupled with the FLEX positioner’s 2.5x patrol radii. The fiber layout sequence in the mirrored module, in this case the LH module, is offset to accommodate the parallelogram “lean” to preserve the pattern.

\FloatBarrier

\begin{figure}[ht]
    \centering
    \includegraphics[width=1\linewidth]{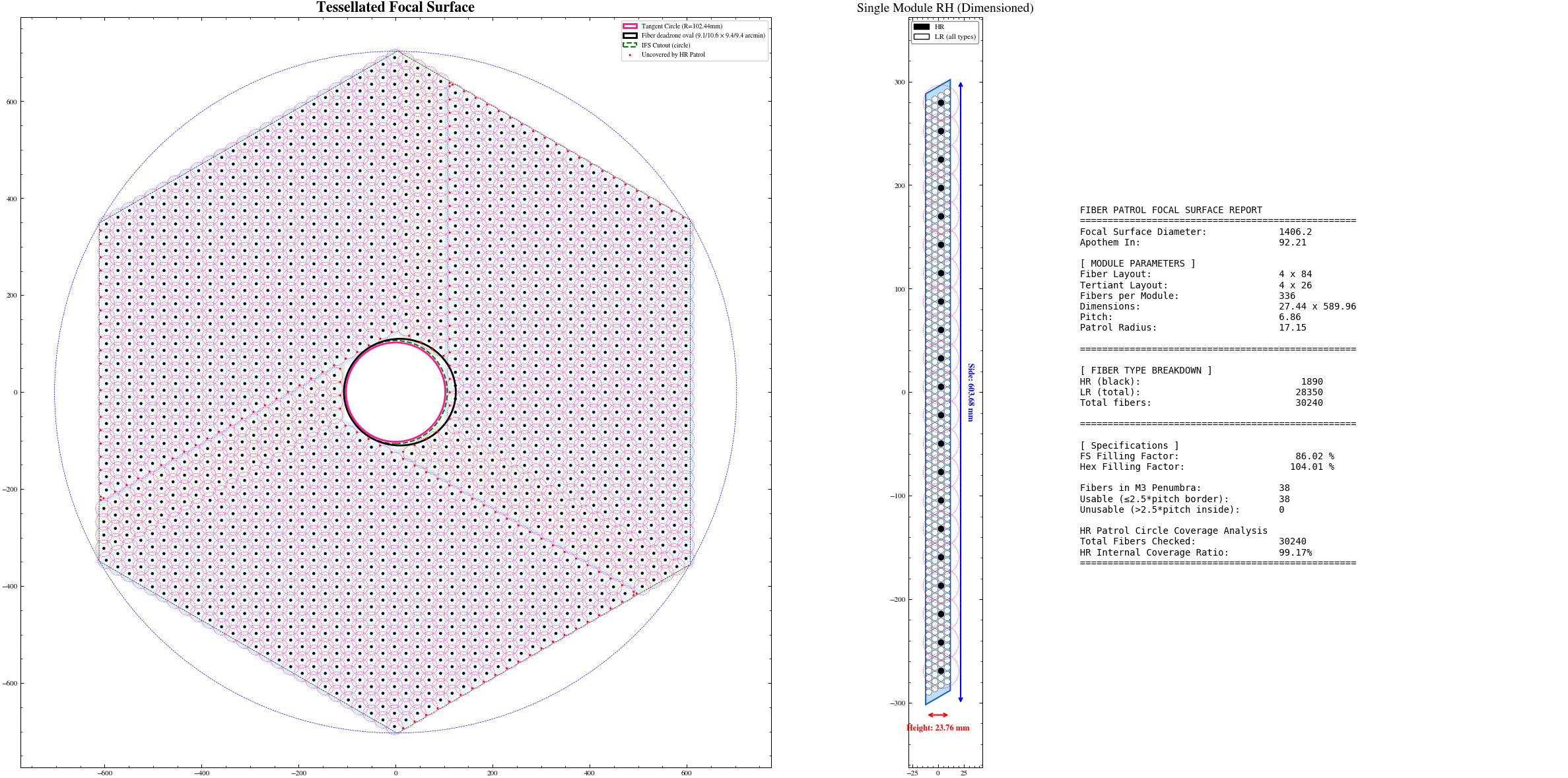}
    \caption{Top-view of fully tiled High Resolution fiber focal surface coverage}
    \label{fig:focal plane - HR}
\end{figure}

Coupled with the fiber layout configuration seen in Figures \ref{fig:focal surface - zoomed}\&\ref{fig:module - HR LR}, the fully tessellated focal surface layout
provides a HR internal coverage of 99.17\%, seen in Figure \ref{fig:focal plane - HR}.

\FloatBarrier

In 3 dimensions, the module sides are great circles, while the ends form circles of latitude. The width of each module tapers by 0.7\% (widest at the inner end). The 4 rows of positioners are parallel, with the pitch preserved across adjacent modules at their outer ends. There is progressively more clearance between positioners in adjacent modules towards their inner ends. The outer hexagonal shape is a mixture, but the difference from a great circle is only $\sim$1~mm, easily taken up in the outer hexagonal frame.  

\begin{figure}[ht]
    \centering
        \hfill
    \begin{subfigure}[b]{0.48\textwidth}
        \centering
        \includegraphics[width=\textwidth]{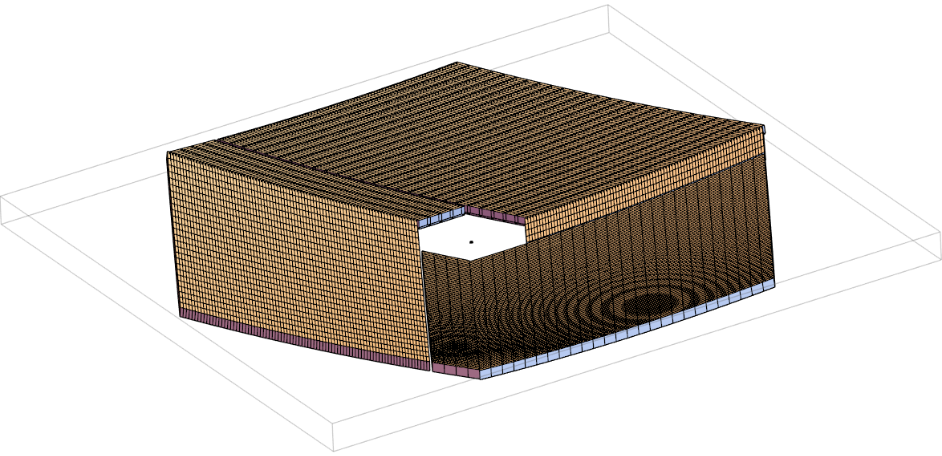} 
        \caption{Isometric view showing the modular tessellation of the parallelogram units across the curved focal surface.}
        \label{fig:3D_offset_view}
    \end{subfigure} 
        \hfill
    \begin{subfigure}[b]{0.48\textwidth}   
        \centering
        \includegraphics[width=\textwidth]{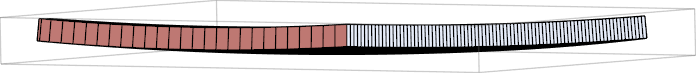} 
        \caption{Side view highlighting the spherical profile of the focal surface and the radial alignment of the assemblies.}
        \label{fig:3D_side_view}
    \end{subfigure}

    \vspace{10pt}
    \caption{Three-dimensional representation of the fully populated WST focal surface. The isometric view (a) highlights the uniform tiling and hexagonal outline, while the side view (b) shows the spherical curvature of the focal surface}
    \label{fig:3D_combined_analysis}
\end{figure}

\begin{figure}[ht]
    \centering
    \includegraphics[width=0.85\linewidth]{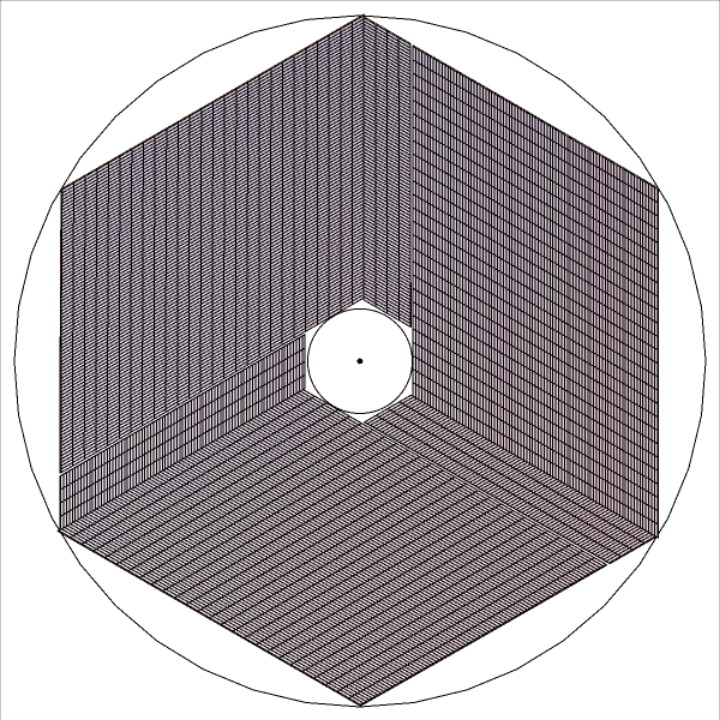}
    \caption{Top view of complete layout of the curved modules. Each parallelogram contains 4 FLEXs. Circles have radius 703mm and 106.5mm}
    \label{fig:top_view_focal_surface}
\end{figure}

The module edges are great circles. But the sides of each strut is a great circle on one side, vs a circle of latitude on the other. The difference amounts to $\sim$60~$\mu$m along the length of the strut, easily taken up within its $\sim$6~mm width. To preserve pitch spacing across module neighbors, the modules are tiled seamlessly together across the trients. The 3-dimensional fully tiled focal surface can be seen in Figures \ref{fig:3D_combined_analysis} \& \ref{fig:top_view_focal_surface}.

\FloatBarrier

\FloatBarrier

\subsection{Summary \& Future Work}

In summary, the modified triskele tiling architecture satisfies the mechanical, optical, and operational constraints imposed by the WST design. The layout accommodates the central IFS pickoff mirror through an asymmetrical distribution of 26 right-handed and 4 left-handed modules per trient, balancing structural symmetry with the preservation of hexagonal packing across modules for optimal fiber routing logic. When paired with the extended $2.5\times$ pitch patrol radius of the FLEX positioner, this layout achieves 99.17\% internal coverage for high-resolution observations. The resulting curvilinear modules tile together seamlessly to map 30,240 active fibers across the hexagonal field-of-view, providing a highly scalable and robust baseline for the next phase of the telescope's instrument development. The summarized geometric and fiber distribution parameters can be seen in Table \ref{tab:focal_surface_summary}.

\begin{table}[h]
\centering
\begin{tabular}{lr}
\toprule
\textbf{Parameter} & \textbf{Value} \\
\midrule
\textit{Focal Surface} & \\
\midrule
Focal Surface Diameter & 1406.2 mm \\
Inner Cutout Diameter & 184.4 mm \\
\midrule
\textit{Module Parameters} & \\
\midrule
Fiber Layout & $4 \times 84$ \\
Trient Layout & $4 \times 26$ \\
Fibers per Module & 336 \\
Pitch & 6.86 mm \\
Patrol Radius & 17.15 mm \\
\midrule
\textit{Fiber Type Distribution} & \\
\midrule
HR Fibers & 1890 \\
LR Fibers (total) & 28350 \\
Total Fibers & 30240 \\
\midrule
\textit{Fiber Patrol Coverage} & \\
\midrule
FS Filling Factor & 86.02\% \\
Hexagonal Filling Factor & 104.01\% \\
HR Internal Coverage Ratio & 99.17\% \\
\bottomrule
\end{tabular}
\caption{Focal surface and module parameters summary.}
\label{tab:focal_surface_summary}
\end{table}
\FloatBarrier

Future work will focus on advancing the CAD of the modular focal surface architecture, transitioning from conceptual layouts to detailed mechanical manufacturing designs. Primary emphasis will be placed on the development of the structural mounting interfaces for the curvilinear modules and the optimization of the central Y-strut support assemblies. Additionally, the issues of integration of localized electronics enclosures, cable management paths for high-density fiber routing, and the structural interfaces required to couple the individual robotic positioners with the module chassis, must all be addressed. To ensure structural integrity under environmental loads, comprehensive analysis of thermal expansion profiles and modal frequency resonance studies must also be conducted. These developments will serve to establish a fully verified, instrument-ready optomechanical envelope for the WST focal plane.

\acknowledgments
\noindent This research was supported by the Leibniz Competition grant K593/2024.

\noindent This research was funded by the Australian Government through the Australian Research Council and their Discovery Project Scheme (DP250103698).

\noindent This project has received funding from the European Union Horizon Europe Research and Innovation Action under grant agreement no. 101183153 -WST. Views and opinions expressed are however those of the author(s) only and do not necessarily reflect those of the European Union or the European Research Executive Agency (REA). Neither the European Union nor the REA can be held responsible for them.
\bibliography{references} 
\bibliographystyle{spiebib} 

\end{document}